\pdfoutput=1

\documentclass[11pt]{article}

\usepackage[final]{ACL2023}

\usepackage{times}
\usepackage{latexsym}

\usepackage[T1]{fontenc}

\usepackage[utf8]{inputenc}

\usepackage{microtype}

\usepackage{inconsolata}

\usepackage{graphicx}

\usepackage{multirow}

\title{Threefold model for AI Readiness: A Case Study with Finnish Healthcare SMEs}

\author{Mohammed Alnajjar \\
  Sparkka Oy \\
  Vantaa, Finland \\
  \texttt{firstname@sparkka.com} \\\And
  Khalid Alnajjar \\
  Rootroo Ltd \\
  Helsinki, Finland \\
  \texttt{firstname@rootroo.com}\\\And
  Mika Hämäläinen \\
  Metropolia University \\ of Applied Sciences \\
  Helsinki, Finland \\
  \texttt{first.lastname@metropolia.fi} \\}

\begin{document}
\maketitle

\begin{abstract}
This study examines AI adoption among Finnish healthcare SMEs through semi-structured interviews with six health-tech companies. We identify three AI engagement categories: AI-curious (exploring AI), AI-embracing (integrating AI), and AI-catering (providing AI solutions). Our proposed threefold model highlights key adoption barriers, including regulatory complexities, technical expertise gaps, and financial constraints. While SMEs recognize AI’s potential, most remain in early adoption stages. We provide actionable recommendations to accelerate AI integration, focusing on regulatory reforms, talent development, and inter-company collaboration, offering valuable insights for healthcare organizations, policymakers, and researchers.
\end{abstract}

\section{Introduction}
The healthcare industry spans multiple sectors, including pharmaceuticals, diagnostics, medical procedures, and wellbeing services. Artificial Intelligence (AI) has demonstrated significant potential in transforming healthcare by assisting in tasks such as medical imaging \cite{suzuki2017overview}, speech processing \cite{partanen2020speech}, and personalized treatment plans \cite{vahedifard2023practical}. Recent advances in neural models have further enhanced AI’s ability to improve diagnostics and patient care while reducing the workload on healthcare professionals (see \citealt{javaid2023chatgpt}).

Health-tech companies play a crucial role in AI-driven innovation, continuously developing new tools and services to enhance medical outcomes and operational efficiency. Governments and private enterprises invest heavily in AI-driven medical research, yet adoption remains complex due to challenges such as regulatory restrictions (e.g., GDPR), data privacy concerns (c.f. \citealt{hamalainen2024legal}), and the risks of computational errors in diagnosis (see \citealt{dave2023chatgpt}).

One notable example of AI’s impact in healthcare is Google's model \citep{nabulsi2021deep}, which demonstrated high sensitivity in detecting abnormal chest conditions, including COVID-19, from X-ray scans. The model's success, despite not being trained on COVID-specific data, highlights AI’s potential in identifying unseen diseases—an essential feature for future medical advancements.

While Finnish health-tech companies acknowledge AI’s transformative potential, their adoption remains limited due to data accessibility, compliance barriers, and the need for extensive validation. This study investigates how Finnish SMEs in healthcare integrate AI into their operations, the challenges they face, and pathways to overcoming these obstacles. We introduce a threefold model categorizing AI adoption among SMEs and provide actionable recommendations to support AI integration in healthcare settings.

AI in healthcare extends beyond clinical applications into digital humanities, where natural language processing (NLP) plays a crucial role in analyzing medical texts, patient records, and healthcare policies. Understanding AI adoption in healthcare SMEs contributes to the broader discourse on how AI, NLP, and computational tools shape interdisciplinary research and real-world applications. 

Our main contributions in this paper are:

\begin{itemize}
    \item Conduct interviews with small and medium-sized healthcare enterprises to assess AI adoption challenges and opportunities.
    \item Perform qualitative analysis to evaluate AI maturity levels among Finnish SMEs.
    \item Introduce a threefold model of AI adoption in business.
    \item Provide strategic recommendations to enhance AI utilization in healthcare.
\end{itemize}

By bridging AI research and practical healthcare applications, this work contributes to the ongoing dialogue on AI's role in healthcare, policy, and digital humanities, offering insights for both researchers and industry practitioners.

\section{Background}
In recent years, artificial intelligence (AI) has significantly impacted various industries, and the healthcare industry is no exception. \citet{reddy2019artificial} explored the incorporation of AI in healthcare delivery, identifying challenges and opportunities for large-scale use while addressing issues such as medical responsibilities and data access. They utilized a qualitative method based on observations of existing AI technologies and predictions of future developments.

In another research, \citet{garbuio2019artificial} analyzed the complexity of value-users in healthcare and emerging business models in AI-driven healthcare startups. By examining archetypes of business models used by entrepreneurs worldwide, they conducted a quantitative analysis of 30 healthcare startups that deploy AI. They concluded that designing effective business models is crucial for bringing beneficial technologies to the market.

AI definitions and deployment status for medium-sized companies were investigated by \citet{ulrich2021relevance}, particularly German SMEs, and the opportunities of AI in supply chain optimization. They collected both quantitative and qualitative data through an open and closed survey questionnaire from 12,360 German companies' emails via the Nexis database. Their findings highlighted the relevance of technologies for companies, AI opportunities in SMEs, and barriers to AI adoption.

The research conducted by \citet{bettoni2021ai} focused on the challenges of applying AI in companies and AI maturity models. They conducted face-to-face interviews and reviewed state-of-the-art literature, examining two SMEs in Poland and Italy. Their research resulted in a conceptual framework to support AI adoption in SMEs.

\citet{bunte2021hard} studied the application of AI in manufacturing and its utilization in the industrial environment, particularly in measuring the financial impact of AI. They employed a mixed-methods approach, using open-ended online questionnaires to collect data from 441 participants across 68 companies in Germany, Austria, and Switzerland. Their research suggested potential strategies to support AI usage in SMEs and identified two best practice solutions.

\section{Methodology and Data}
This section explains the research strategy and data collection methods used. This work employs a case study research strategy, focusing on health-tech companies in Finland as the unit of analysis. The case study methodology allows us to explore complex phenomena and gain insights into the underlying dynamics and mechanisms that drive them \citep{yin2009case}. Through semi-structured interviews \citep{saunders2009research}, rich and detailed data from various stakeholders can be collected within the health-tech industry.

To apply the case study methodology in this research, first, a literature review was conducted to identify relevant theories and concepts that would inform the research questions. Then, data was collected through semi-structured interview methods. The data were analyzed using a qualitative data analysis software, following a systematic approach. The results of the analysis informed discussions and recommendations for further research and practice.

The research methodology consists of qualitative approaches that include four parts: the first is researching the existing healthcare and wellbeing SMEs in Finland to analyze their products and services offerings to build a general understanding of health-technology applications in the market. Then, in the second phase, the companies that have digital products or services have been contacted to request interviews with them. The third part involves analyzing the findings from the conducted interviews with health-tech Finnish companies about their AI usage level with a focus on Small- and Medium-sized Enterprises (SMEs). Finally, the development section provides insights and recommendations to support the use of AI in Finnish health-tech businesses, based on the interviews conducted, and an overview of the future possible uses of AI in the sector.

\section{Results}
In this section, the results of interview analysis reports will be presented comprehensively, in a scientific manner that allows understanding and analysis of the results. As well, the most important phrases that were mentioned in the interviews about the research topic. In addition, the common observations, and trends related to the use, benefits, and challenges of embedding AI in the healthcare and wellbeing sector.

The key findings of the results are based on the analysis of the data collected through the interviews with the respondents. Appendix 1 includes some of the respondents' citations to provide further insight into the themes that emerged. However, not all key findings are represented in the appendix, as some were not explicitly stated by the respondents but were inferred from the overall interview. Therefore, it is important to read the entire interview transcript to fully understand the key findings.

This section is divided into subsections according to the key findings categories. In each subsection, it will cover the main findings by dividing the answers into groups. Also, a synthesis for the findings for each question is provided.

\subsection{AI in Products and Services}
Companies have considered using AI or are already embracing it in very different ways. Based on the interviews, it's possible to identify three different ways companies are using or considering AI in the health care sector. AI can be used as a tool for analysis, seen as a possible future solution or provided as their core product.

Companies 2, 3 and 5 reported that they use AI to conduct analysis on health data. Company 5 used AI to automatically detect anomalies in ECG analysis results whereas the other two companies used AI in a less autonomous way to analyse data for medical professionals to make better judgments. Company 2 believed firmly that they could automatize even the step where a medical professional needs to take a look at the results and have an AI diagnose and interpret the data as well.

Companies 1 and 6 are looking into using AI in their work. Company 6 has identified that their problem of working with brain data related to epilepsy is often predictable. They envision embracing AI in the future to automatically identify when an epileptic seizure is about to happen. Company 1 has taken some steps towards processing fundus images automatically by shortlisting potential companies whose technology is mature enough to detect anomalies in such data. However, Company 1 points out the issue arising from a limited amount of data for training AI models, which might make their AI aspirations unfeasible.

Company 4 stands out from the crowd by being the only company that provides AI services as their main product. They are primarily a machine learning company and their task is to cater for health AI related needs of their clients. They provide AI solutions for diagnostic needs.

\subsection{How AI is Defined}
AI is quite a flexible notion as it can consist of many different aspects of computing starting from simple programming to machine learning. This section will describe how the interviewed companies understand the word AI. The companies defined AI as a learning algorithm, quality of life enhancer and human-level anomaly detector.

Most of the companies (1, 2, 3 and 4) had a modern definition for AI, that is that it is some sort of an algorithm that ends up learning predictions based on data. Company 2 highlighted the importance of speed and that AI can be used to partially replace a costly medical specialist, however, they pointed out that medical doctors do not easily accept their AI colleague but refer to issues like privacy concerns. Company 3 also pointed out the problem of privacy by mentioning EU regulations on the use of medical data. Company 1 defined AI narrowly from the point of view of a learning classifier. They saw the lack of clean training data as an issue and a hindrance in developing AI. Company 4 wanted to point out that AI is such a large field that from their point of view, they are dealing with machine learning rather than AI.

Company 6 had not yet embraced AI, which is something reflected in the way they understood AI. For them it is a question of a quality-of-life improvement over not using an AI at all. Company 5 had the highest hopes for AI by defining it as a human-level anomaly detector. This answer differs from the majority in the sense that the company sees AI as an unsupervised tool that can detect tendencies from data without being an actively learning agent.

\subsection{Perceived Level of AI Maturity}
This section will describe how companies perceived their own level of AI maturity following the levels established by Gartner. The interviewed companies self-identified as being in categories 1, 2 and 3. These are well in line with the discussion in the earlier sections which means that their self-reporting is rather honest in terms of how they described they actually used AI tools.

Companies 1 and 6 reported level 1 as their own level. Company 6 highlighted the issue of costs related to transitioning from an AI-aware level into a level where AI is actively used. There are costs not only related to development but also related to conforming with all regulations that are in place. Company 1 reported that their own level is currently 1, but they estimated the level of their short-listed future collaborators to be 3.

Companies 2, 3 and 5 reported their level to be 2, that is the level in which AI is applied mostly for data science needs. Company 2 also identified that they are envisioning a medical head instrument that is currently on the level 1 of AI maturity.

Company 4, which is the one relying solely on AI in their business model, was the only one reporting their level to be as high as 3. This is the level of AI in production where new value is being created through AI. The company does not have any aspirations to climb higher on the AI maturity levels because they are a small company and cannot reach the stars.

\subsection{AI Application Areas}
This section will describe how the interviewed companies for this study use AI outside of the main application area that has been described in the earlier sections. Mostly none of the companies really uses AI for any other business applications. Companies 1, 3, 4 and 6 failed to give any example on how they would utilize AI in other areas.

Company 2 identified that they do use AI in marketing. They host an AI-powered chatbot on their website. Apart from this, the company did not identify other uses for AI in their business.

Company 5 pointed out an unintentional use of AI. They only use AI in other areas because it is already baked in the software they use on a daily basis such as Microsoft and Atlassian tools.

These insights provide an overview of how Finnish SMEs in the health-tech sector are utilizing AI and their perspective on its maturity and application areas. The next sections will continue discussing the challenges faced and the perceived impact of AI adoption in the healthcare industry.

\subsection{Perceived Impact of AI}
AI is hardly used just because it is trendy but because it has a tangible impact on how business is conducted. This section describes what the interviewed companies had to say about the impact AI has had on their work. The interviewed companies thought rather unanimously that AI is indispensable for their operations. Only Company 6 reported that AI had no impact thus far, but this was due to the fact that the company had not started to use AI yet. Interestingly, even Company 1, which does not yet use AI, reported that AI is a must-have, which explains why they are actively seeking a suitable AI collaborator.

Companies 1, 2, 3, 4, and 5 stated that AI is essential. Company 2 identified that conducting the level of analysis they need to do would be impossible without AI methods. Company 4, which is purely an AI-based company, stated that they would not have any market value without AI. Furthermore, Company 4 indicated that embracing AI gave them an advantage in acquiring funding.

\subsection{Data Source}
It is no secret that AI relies heavily on data. Just as the definitions for AI suggested by the interviewed companies, modern AI is mainly about learning from data. This section describes the findings on what data sources the companies rely on. Data is either collected in-house or obtained from external providers.

Companies 2, 4, 5, and 6 report using in-house data. In the case of Company 6, it is stated as a possible hypothetical data source. For other companies, they report that their data comes from different measuring devices that monitor patients, such as ECG and EEG. The aforementioned companies have not considered the need for additional complementary data from other companies or open repositories.

Companies 1 and 3 use external providers. Company 1 stated that they collaborate with manufacturers of different fundus cameras to gain more data. Company 3 has access to big data; however, they are still looking for ways to benefit from it. This is understandable given that big data may conceal answers to many questions one does not even think of initially.

\subsection{Computing Environment}
Given that AI relies heavily on data, another issue needs to be taken care of: the computing environment. AI models need to be trained on data, which might require high usage of computational resources. This section describes the computing environments the companies used. The companies had either outsourced AI tools entirely, used a private server, or a public cloud.

Companies 1, 2, and 6 stated that either their AI tools are provided by third parties or that they will be provided by third parties. Company 2 further mentioned that their team is too small to handle their own computing environment for AI needs.

Companies 4 and 5 use public cloud providers, Amazon AWS and Microsoft Azure, respectively. Company 3 uses public clouds for training AI models and private servers to handle personal data. Company 6 envisions that they will start with a private server and, if needed, move to a public cloud.

\subsection{Challenges in AI}
New technology might seemingly come with all the bells and whistles, but embracing it is not always straightforward. This section describes the challenges the informants faced when implementing AI and when using it. These challenges can be categorized into three main areas: regulations, market acceptance, and talent acquisition.

If all the companies were interviewed simultaneously, they would likely have said in unison that the EU has regulations that are too strict for health-related AI. Companies 2, 3, 4, and 6 all stated that the USA has more lenient laws on many aspects. Companies 2 and 4 had issues with personal data regulations in the EU. Additionally, Company 4 mentioned facing legal challenges when trying to get approval for their technology. Company 6 faced issues with strict medical certification requirements that, again, are more relaxed in the USA and China. Finally, Company 3 mentioned that the US medical authority FDA allows the use of AI models that are continuously learning from data whereas the EU allows models that are trained once and tested on at least 200,000 samples. Thus, their challenge was related to the inflexibility of the regulations.

Companies 1 and 2 had issues with market acceptance. Both companies reported that it was difficult to get approval from medical professionals on the customer side to start using the AI in production. New technology is often met with a degree of resistance and skepticism, which might explain these findings.

Companies 5 and 6 reported a more concrete issue of being able to find competent members of staff. Both companies struggle to find people with a suitable medical background and a necessary set of R\&D skills in the field of AI. Perhaps this is explained by the fact that machine learning and medicine are taught as very different subjects in many universities with little to no overlap.

\subsection{Perceived Benefits of AI}
In terms of benefits, the interviewed companies saw two main advantages: speed and accuracy, and indispensability. This section briefly describes what the companies had to say about these, although there are probably many more benefits that the informants did not consider during the interview.

Companies 2, 3, 5, and 6 stated that the main benefit of AI is that it can perform laborious analysis work faster than a human being and do so with high accuracy. This means that the problems the companies deal with are also defined well enough that the AI models have learned not to err frequently.

Companies 1 and 4 continued to see AI as a necessity. In the case of Company 4, AI truly is their lifeline given that their entire operations revolve around providing AI services. Company 1 also stated that there is a lot of room in the market and a lot of unsatisfied innovation potential, especially in the EU for health-related AI tools, unlike in Asia, where the market is already oversaturated.

\subsection{Wishes for Third Parties}
This section describes what needs the companies reported they would have for third-party services to support their AI ventures. Interestingly, Companies 4, 5, and 6 reported absolutely nothing. For the other companies, the needs can be classified into access to resources and budget solutions.

Companies 1 and 2 stated that they would be interested in having access to more data from external providers. Given that AI runs on data, it is no surprise that such a need might emerge. As described in earlier sections, many companies relied heavily on their in-house data, but even so, in the world of AI, more is always better.

Companies 2 and 3 also expressed a need for low-cost access to AI. Especially Company 2 stated that the typical price tag of €300,000-€400,000 for an AI project developed by an external company is way too high for a small business. Company 2 suggested either lower prices or access to funding as a solution. Company 3 advocated for cheaper access to high-performance computing so that AI models can be trained in a more cost-efficient manner.

\subsection{Future Concerns}
The field of AI is currently in an ever-changing state with continuous innovations taking place in all areas of AI. This section describes how the informants see what the future holds for their companies in relation to AI. The interviewed companies had many different ideas for the future: solutions for staff shortages, positive changes in healthcare, use in other business aspects, changes in regulations, better AI models, and higher computational requirements.

Company 2 sees AI as one possible solution for staff shortages that result from a variety of factors such as an aging population. They also believe that AI will bring a positive change to how healthcare services are provided. For example, a patient would not need to wait a long time to see a neurologist if an AI model could diagnose the symptoms automatically.

Company 4 foresees a clear regulatory need for introducing standards to healthcare data and AI models. The current situation is a Wild West with no cohesive practices. Meanwhile, Company 1 presents a practical issue still challenging for modern AI techniques: detecting more than one symptom at a time accurately.

\subsection{Advice for Other Companies}

When the companies were asked about the advice they would give to another company that has not yet considered AI at all, they provided responses that fit into the following categories: gathering data, defining the problem, hiring competent people, and starting experimentation. These steps, combined, form a practical roadmap for companies looking to integrate AI into their business operations.

Companies 2 and 3 emphasized the importance of \textit{gathering data}. As Company 2 puts it, it is better to start collecting data sooner rather than later, even if AI plans are not in the near future. Data is the foundation of AI, and having access to well-structured datasets ensures a smoother transition when the company is ready to implement AI solutions. Company 3 also noted that it is important to analyze the collected data to understand what value can be extracted from it.

Companies 1 and 2 discussed the significance of \textit{defining the problem}. The sooner a company has clarity regarding the problem it wants to solve, the sooner it will know what type of data is required, according to Company 2. Company 1 pointed out that taking extra care in specifying goals correctly from the beginning is essential, as unclear objectives can lead to inefficient AI implementation and wasted resources.

Company 6 recommended that businesses \textit{hire competent people} with the right mix of skills. They highlighted the challenge of finding professionals who possess both AI expertise and knowledge in the healthcare sector. The integration of AI in healthcare requires interdisciplinary collaboration between AI experts, medical professionals, and business strategists.

Companies 4 and 5 encouraged businesses to \textit{start experimenting} with AI. They emphasized the need for companies to test AI solutions in small, controlled environments before fully committing to large-scale implementations. Company 5 suggested that businesses should begin by playing around with their data to gain a deeper understanding of potential AI applications. Company 4 also pointed out that many AI tools and frameworks are readily available, making it easier for businesses to start experimenting with AI-driven solutions.

\section{The Threefold Model of AI in Business}
To complete the development task of this research, a collaborative brainstorming development method was utilized. This method involves generating a large number of ideas and then selecting the most promising ones to pursue further \citep{wilson2013brainstorming}. A brainstorming session with two members from the commissioner was organized to generate solutions on what services can pave the way for health-tech SMEs to adopt and develop the use of AI, and how health-tech companies can cooperate together to elevate the level of AI in the sector. The ideas were then grouped and analyzed to identify the most relevant and feasible ones. This collaborative method allowed the identification of potential gaps in AI utilization in the health-tech sector and to come up with a state-of-the-art framework.

Based on the findings during the interviews, a threefold model on the use of AI in business has been elaborated. The following three categories have been identified for AI in business: AI Curious, AI Embracing, and AI Catering companies. This section will shed more light on each of these categories and how they differ from each other. The categorization is based on how AI is operationalized in different companies.

The model is useful when trying to understand and better analyze the use of AI from a grassroots level. This can help companies better locate themselves in terms of AI and business. One company does not need to fit in only one of the categories either, but a company can, for example, be AI Catering in providing a specific solution for healthcare and AI Curious when planning on integrating AI into marketing practices.

It is important to note that the highest levels of AI maturity are not part of this framework because the interviewed companies would not place themselves that high in the hierarchy. This tells us also something about the paradigm shift in the field of AI, where hardcore AI research and development is in the hands of larger companies such as Google, Meta, or OpenAI, while the field-specific use of AI is often handled by companies that do not have massive resources for core AI research. Revolutionary AI methods such as word embeddings \citep{mikolov2013efficient} and the Transformer model \citep{vaswani2017attention} have been developed by Google, models such as ChatGPT and DALL-E \citep{ramesh2022hierarchical} by OpenAI, and audio embeddings \citep{baevski2020wav2vec} by Meta. In short, there is no room for a small player to compete in the space of new AI revolutions.

\begin{figure}[t]
    \centering
    \includegraphics[width=\columnwidth]{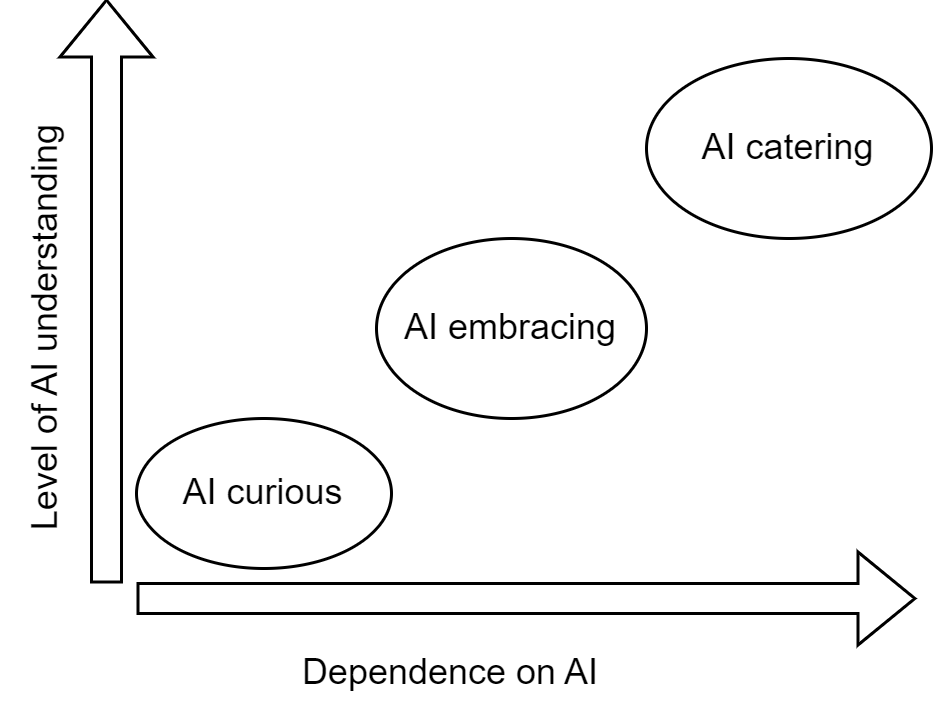}
    \caption{Threefold Model of AI in Business}
    \label{fig:threefold}
\end{figure}

\subsection{AI Curious}
An AI Curious company is still in the process of planning. Such a company can be currently identifying possible problems where AI can be used or can already be in talks with AI-providing companies about solving a particular problem. AI curious companies may engage in practices of collecting data and analysing it to uncover its potential in the future when the company is ready to start using AI in their day-to-day operations.

AI curiosity can thus be a time of great exploration of both AI and its added value to the target market. AI curious companies could benefit from cost-efficient consultants, external R\&D funding, and existing tools and datasets. The first stop for an AI curious company might thus be an open data platform such as Zenodo or Kaggle or an open AI model platform such as Huggingface or Model Zoo.

At this stage, it is important that the company has a clear idea of what the actual AI problem is before moving to the next stage of embracing AI. Moving forward with an ill-defined problem might have costly consequences or poor market adaptation. This calls for a degree of understanding of the current limitations and possibilities of what can and cannot be done with AI. This understanding can be acquired internally or externally.

\subsection{AI Embracing}
AI Embracing companies have already started to use AI in their business operations. However, they either do not develop AI by themselves but buy it as a service from an external provider or if they develop AI in-house, it is not their main product but rather an auxiliary tool for their actual product that could be provided without AI as well.

An AI Embracing company has identified one or a few targeted problems that they can optimize with AI. They, however, are not fully relying on AI because AI is used as a functional part of their business pipeline that also consists of manual tasks such as the final analysis or diagnosis of the numbers crunched by AI.

A strong collaboration between an AI Embracing company and their AI provider is advised. Modern AI is entirely data-driven, and thus better results can be obtained if the AI Embracing company is capable and willing to share their own data with their AI provider. An AI Embracing company might run AI models on their own servers or on an external cloud over an API access.

\subsection{AI Catering}
Companies that are AI Catering provide AI services to other companies that are currently only embracing AI or in the AI Curious stage. AI products and services are the core offerings of AI Catering companies. AI Catering companies do not necessarily develop their own cutting-edge AI solutions, but they can rather use existing AI methods, such as Transformers \citep{wolf2019huggingface}, Datasets \citep{lhoest2021datasets}, PyHFST \citep{alnajjar2023pyhfst}, Gensim \citep{rehurek}, and SciKit \citep{pedregosa2011scikit}, that they train on in-domain data to solve the business problems their customers have.

AI Catering companies can provide and train AI models on their own servers or outsource the heavy computation to a cloud provider such as AWS or Azure. While AI is typically provided as a service, AI Catering companies may provide their solutions so that their clients can run the AI models on their own machines.

Because real state-of-the-art AI development has moved beyond the reach of smaller companies, AI Catering companies can only truly compete against each other with data. The more and better-quality data an AI Catering company has, the better their AI models will be and the more advantage they will have in the market. Access to computational resources plays an important role here as well. Large amounts of data require more computational power to be harnessed in use.

This threefold model provides a structured way to understand AI adoption in SMEs, helping businesses navigate their AI journey more effectively. The next section will discuss inter-categorical business opportunities and how companies within these three categories can collaborate to maximize the benefits of AI.

\begin{figure}[h]
    \centering
    \includegraphics[width=\columnwidth]{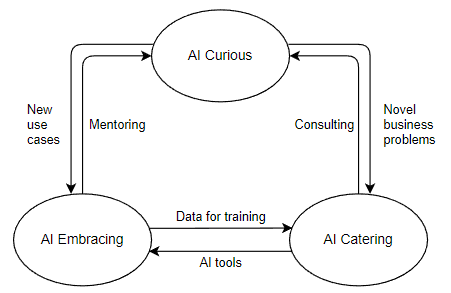}
    \caption{Interdependence of Companies in Different Categories of Business AI}
    \label{fig:interdependence}
\end{figure}

\section{Discussion}

Most companies are still in the early stages of AI adoption, either experimenting or in the initial implementation phase, with none reaching high maturity yet. Despite recognizing AI’s potential to improve healthcare, they face three major challenges: legal, technical, and financial.

All interviewees highlighted that regulatory compliance is the most significant barrier. The legal framework governing AI in healthcare has not kept pace with technological advancements, creating hurdles for innovation \citep{powles2017google}. In Finland and the EU, medical licensing is a complex, evolving process, making AI implementation difficult \citep{petersson2021challenges}. For instance, conducting clinical trials for a new AI-powered medical device requires national authority approval, which can take months and necessitate \textbf{180,000 tests} to ensure safety \citep{jiang2017artificial}. Even after approval, further ethical committee clearance is required, adding time and financial strain.

These findings underscore the urgent need for regulatory improvements to support AI adoption in Finland’s health-tech sector. Streamlining medical licensing and approval processes would reduce delays and costs, allowing companies to focus on innovation and deployment of AI solutions.

Two-thirds of interviewees also cited difficulties in finding qualified AI professionals with both medical expertise and programming skills. \textit{"There are currently few AI experts in the health-tech industry,"} stated one company leader. Other technical barriers include collecting and cleaning reliable data, integrating AI with hospital systems, and overcoming resistance from medical professionals, who often require years to accept new technologies.

Financial constraints further limit AI adoption. High upfront costs, delayed return on investment, limited access to training data, and expenses related to regulatory compliance present significant barriers. Privacy restrictions and legal requirements further complicate AI implementation for SMEs in the sector.

Based on interviews and analysis, this study recommends several measures to enhance AI utilization in health-tech SMEs, particularly in Finland and the EU. Regulatory frameworks should be reformed to facilitate AI integration while ensuring data privacy and patient safety. Simplifying research processes, streamlining medical licensing, and reducing bureaucratic barriers will allow companies to focus on innovation rather than compliance hurdles.

Developing AI talent is crucial, as many SMEs struggle to find qualified professionals. A solution is to invest in continuous training programs tailored to healthcare AI, lowering entry barriers while enhancing expertise. Additionally, attracting skilled foreign professionals with competitive salaries and benefits can bridge the talent gap.

Financial barriers remain a major obstacle for AI adoption, given the high initial investment and delayed returns. Establishing funding mechanisms, including government grants and private investments, would provide SMEs with the resources needed to integrate AI solutions. Collaboration platforms can further support AI adoption by connecting SMEs with research institutions, AI service providers, and industry partners, enabling knowledge-sharing and joint innovation.

Ensuring data security is another priority. A certification system for AI data privacy should be introduced, verifying that companies handling sensitive medical data comply with strict security standards. This would enforce encrypted storage, controlled access, and detailed audit logs to safeguard patient information while enabling responsible AI deployment.

These recommendations offer practical strategies for policymakers and industry stakeholders to foster AI adoption in the Finnish health-tech sector, balancing innovation with regulatory compliance and data security.

\section{Conclusion}

This study examined AI adoption among Finnish health-tech SMEs, identifying key challenges and opportunities. While AI holds immense potential for enhancing healthcare efficiency and patient outcomes, most SMEs remain in early adoption stages due to regulatory barriers, limited AI expertise, and financial constraints. A more flexible legal framework, improved access to AI talent, and increased funding opportunities are necessary to accelerate AI integration in healthcare.

Addressing these challenges will enable SMEs to leverage AI for innovation, ultimately benefiting the healthcare sector and society. Future work should explore collaborative AI development models, interdisciplinary training programs, and policy reforms to foster AI adoption. By streamlining regulations and promoting industry partnerships, AI-driven solutions can be more effectively implemented, ensuring sustainable growth and improved patient care.

\bibliography{anthology,custom}

\begin{thebibliography}{27}
\expandafter\ifx\csname natexlab\endcsname\relax\def\natexlab#1{#1}\fi

\bibitem[{Alnajjar and H{\"a}m{\"a}l{\"a}inen(2023)}]{alnajjar2023pyhfst}
Khalid Alnajjar and Mika H{\"a}m{\"a}l{\"a}inen. 2023.
\newblock Pyhfst: A pure python implementation of hfst.
\newblock In \emph{Lightning Proceedings of NLP4DH and IWCLUL 2023}, pages 32--35.

\bibitem[{Baevski et~al.(2020)Baevski, Zhou, Mohamed, and Auli}]{baevski2020wav2vec}
A.~Baevski, H.~Zhou, A.~Mohamed, and M.~Auli. 2020.
\newblock \href {https://github.com/pytorch/fairseq} {wav2vec 2.0: A framework for self-supervised learning of speech representations}.

\bibitem[{Bettoni et~al.(2021)Bettoni, Matteri, Montini, Gladysz, and Carpanzano}]{bettoni2021ai}
A.~Bettoni, D.~Matteri, E.~Montini, B.~Gladysz, and E.~Carpanzano. 2021.
\newblock \href {https://doi.org/10.1016/J.IFACOL.2021.08.082} {An ai adoption model for smes: A conceptual framework}.
\newblock \emph{IFAC-PapersOnLine}, 54(1):702--708.

\bibitem[{Bunte et~al.(2021)Bunte, Richter, and Diovisalvi}]{bunte2021hard}
A.~Bunte, F.~Richter, and R.~Diovisalvi. 2021.
\newblock \href {https://doi.org/10.5220/001020410614062051} {Why it is hard to find ai in smes: A survey from the practice and how to promote it}.
\newblock In \emph{ICAART 2021 - Proceedings of the 13th International Conference on Agents and Artificial Intelligence}, volume~2, pages 614--620.

\bibitem[{Dave et~al.(2023)Dave, Athaluri, and Singh}]{dave2023chatgpt}
Tirth Dave, Sai~Anirudh Athaluri, and Satyam Singh. 2023.
\newblock Chatgpt in medicine: an overview of its applications, advantages, limitations, future prospects, and ethical considerations.
\newblock \emph{Frontiers in artificial intelligence}, 6:1169595.

\bibitem[{Garbuio and Lin(2019)}]{garbuio2019artificial}
M.~Garbuio and N.~Lin. 2019.
\newblock \href {https://doi.org/10.1177/0008125618811931} {Artificial intelligence as a growth engine for health care startups: Emerging business models}.
\newblock \emph{California Management Review}, 61(2):59--83.

\bibitem[{H{\"a}m{\"a}l{\"a}inen(2024)}]{hamalainen2024legal}
Mika H{\"a}m{\"a}l{\"a}inen. 2024.
\newblock Legal and ethical considerations that hinder the use of llms in a finnish institution of higher education.
\newblock In \emph{Proceedings of the Workshop on Legal and Ethical Issues in Human Language Technologies@ LREC-COLING 2024}, pages 24--27.

\bibitem[{Javaid et~al.(2023)Javaid, Haleem, and Singh}]{javaid2023chatgpt}
Mohd Javaid, Abid Haleem, and Ravi~Pratap Singh. 2023.
\newblock Chatgpt for healthcare services: An emerging stage for an innovative perspective.
\newblock \emph{BenchCouncil Transactions on Benchmarks, Standards and Evaluations}, 3(1):100105.

\bibitem[{Jiang et~al.(2017)Jiang, Jiang, Zhi, Dong, Li, Ma, Wang, Dong, Shen, and Wang}]{jiang2017artificial}
F.~Jiang, Y.~Jiang, H.~Zhi, Y.~Dong, H.~Li, S.~Ma, Y.~Wang, Q.~Dong, H.~Shen, and Y.~Wang. 2017.
\newblock \href {https://doi.org/10.1136/SVN-2017-000101} {Artificial intelligence in healthcare: past, present and future}.
\newblock \emph{Stroke and Vascular Neurology}, 2(4):230--243.

\bibitem[{Lhoest et~al.(2021)}]{lhoest2021datasets}
Q.~Lhoest et~al. 2021.
\newblock \href {https://doi.org/10.18653/v1/2021.emnlp-demo.21} {Datasets: A community library for natural language processing}.
\newblock In \emph{EMNLP 2021 - 2021 Conference on Empirical Methods in Natural Language Processing: System Demonstrations}, pages 175--184.

\bibitem[{Mikolov et~al.(2013)Mikolov, Chen, Corrado, and Dean}]{mikolov2013efficient}
T.~Mikolov, K.~Chen, G.~Corrado, and J.~Dean. 2013.
\newblock \href {https://arxiv.org/abs/1301.3781v3} {Efficient estimation of word representations in vector space}.
\newblock In \emph{1st International Conference on Learning Representations, ICLR 2013 - Workshop Track Proceedings}.

\bibitem[{Nabulsi et~al.(2021)}]{nabulsi2021deep}
Z.~Nabulsi et~al. 2021.
\newblock \href {https://doi.org/10.1038/s41598-021-93967-2} {Deep learning for distinguishing normal versus abnormal chest radiographs and generalization to two unseen diseases tuberculosis and covid-19}.
\newblock \emph{Scientific Reports}, 11(1):1--15.

\bibitem[{Partanen et~al.(2020)Partanen, H{\"a}m{\"a}l{\"a}inen, and Klooster}]{partanen2020speech}
Niko Partanen, Mika H{\"a}m{\"a}l{\"a}inen, and Tiina Klooster. 2020.
\newblock Speech recognition for endangered and extinct samoyedic languages.
\newblock \emph{arXiv preprint arXiv:2012.05331}.

\bibitem[{Pedregosa et~al.(2011)}]{pedregosa2011scikit}
F.~Pedregosa et~al. 2011.
\newblock Scikit-learn: Machine learning in python.
\newblock \emph{Journal of Machine Learning Research}, 12:2825--2830.

\bibitem[{Petersson et~al.(2021)Petersson, Larsson, Nygren, Neher, Reed, Tyskbo, and Svedberg}]{petersson2021challenges}
L.~Petersson, I.~Larsson, J.~M. Nygren, M.~Neher, J.~E. Reed, D.~Tyskbo, and P.~Svedberg. 2021.
\newblock \href {https://doi.org/10.1186/s12913-022-08215-8} {Challenges to implementing artificial intelligence in healthcare: a qualitative interview study with healthcare leaders in sweden}.
\newblock \emph{Health Services Research}, 22:850.

\bibitem[{Powles and Hodson(2017)}]{powles2017google}
J.~Powles and H.~Hodson. 2017.
\newblock \href {https://doi.org/10.1007/S12553-017-0179-1} {Google deepmind and healthcare in an age of algorithms}.
\newblock \emph{Health and Technology}, 7(4):351--367.

\bibitem[{Ramesh et~al.(2022)Ramesh, Dhariwal, Nichol, Chu, and Chen}]{ramesh2022hierarchical}
A.~Ramesh, P.~Dhariwal, A.~Nichol, C.~Chu, and M.~Chen. 2022.
\newblock \href {https://arxiv.org/abs/2204.06125v1} {Hierarchical text-conditional image generation with clip latents}.

\bibitem[{Reddy et~al.(2019)Reddy, Fox, and Purohit}]{reddy2019artificial}
S.~Reddy, J.~Fox, and M.~P. Purohit. 2019.
\newblock \href {https://doi.org/10.1177/0141076818815510} {Artificial intelligence-enabled healthcare delivery}.

\bibitem[{Rehurek(n.d.)}]{rehurek}
Rehurek. n.d.
\newblock Gensim--python framework for vector space.

\bibitem[{Saunders et~al.(2009)Saunders, Lewis, Thornhill, Lewis, and Thornhill}]{saunders2009research}
M.~Saunders, P.~Lewis, A.~Thornhill, S.~Lewis, and Thornhill. 2009.
\newblock \href {www.pearsoned.co.uk} {\emph{Research methods for business students fifth edition}}.

\bibitem[{Suzuki(2017)}]{suzuki2017overview}
Kenji Suzuki. 2017.
\newblock Overview of deep learning in medical imaging.
\newblock \emph{Radiological physics and technology}, 10(3):257--273.

\bibitem[{Ulrich and Frank(2021)}]{ulrich2021relevance}
P.~Ulrich and V.~Frank. 2021.
\newblock \href {https://doi.org/10.1016/J.PROCS.2021.08.228} {Relevance and adoption of ai technologies in german smes - results from survey-based research}.
\newblock \emph{Procedia Computer Science}, 192:2152--2159.

\bibitem[{Vahedifard et~al.(2023)Vahedifard, Haghighi, Dave, Tolouei, and Zare}]{vahedifard2023practical}
Farzan Vahedifard, Atieh~Sadeghniiat Haghighi, Tirth Dave, Mohammad Tolouei, and Fateme~Hoshyar Zare. 2023.
\newblock Practical use of chatgpt in psychiatry for treatment plan and psychoeducation.
\newblock \emph{arXiv preprint arXiv:2311.09131}.

\bibitem[{Vaswani et~al.(2017)}]{vaswani2017attention}
A.~Vaswani et~al. 2017.
\newblock Attention is all you need.
\newblock \emph{Advances in Neural Information Processing Systems}, 30.

\bibitem[{Wilson(2013)}]{wilson2013brainstorming}
C.~Wilson. 2013.
\newblock \emph{Brainstorming and beyond: a user-centered design method}.
\newblock Newnes.

\bibitem[{Wolf et~al.(2019)}]{wolf2019huggingface}
T.~Wolf et~al. 2019.
\newblock \href {https://arxiv.org/abs/1910.03771v5} {Huggingface's transformers: State-of-the-art natural language processing}.

\bibitem[{Yin(2009)}]{yin2009case}
R.~K. Yin. 2009.
\newblock \emph{Case study research: Design and methods}, 4 edition.
\newblock Sage Publications.

\end{thebibliography}
\bibliographystyle{acl_natbib}

\appendix
\section{Questions asked in the interview}

Table~\ref{interview-table} lists the questions asked during the interviews.

\begin{table*}
  \centering
  \renewcommand{\arraystretch}{1.3}
  \begin{tabular}{|p{4cm}|p{7cm}|p{6cm}|}
    \hline
    \textbf{Interview Theme} & \textbf{Question} & \textbf{Why was the question asked?} \\
    \hline

    \multirow{4}{4cm}{\textbf{Company’s current position on AI}} 
    & Could you tell us about yourself and your company’s activities? & Answering these questions will contribute to building knowledge about the company’s current position on the use of AI in the healthcare context, thus will contribute to answering RQ1. \\
    \cline{2-3}
    & Do you consider AI use in your business solutions? (\& why?) & \\
    \cline{2-3}
    & At what level of Gartner AI Maturity Model is your company currently?* & \\
    \cline{2-3}
    & What impact has your company experienced from the use of AI? & \\
    \hline

    \multirow{4}{4cm}{\textbf{Challenges associated with the adoption and use of AI in healthcare industry}} 
    & What problems did you face at the initial stage of adopting AI in healthcare? & To explore and search the challenges that Finnish health-tech SMEs face in the adoption and usage of AI, and that is critical to answer RQ2 and to develop real-world practical solutions for them. \\
    \cline{2-3}
    & How did your company resolve these issues? & \\
    \cline{2-3}
    & What are the current challenges the company is facing in applying AI in healthcare? & \\
    \cline{2-3}
    & What future concerns do you expect to exist around the use of AI in healthcare? & \\
    \hline

    \multirow{3}{4cm}{\textbf{Reliable recommendations from the field experts}} 
    & In your opinion, what are the actions that can resolve these challenges? & To collect informative opinions from industry leaders about action plans that can lead to elevate the AI-Maturity level in the sector. \\
    \cline{2-3}
    & What services do you wish to be provided by AI-solution providers to facilitate the emergence of AI among health-tech companies? & \\
    \cline{2-3}
    & What is your advice to start-ups in the health-tech sector on the use of AI? & \\
    \hline
  \end{tabular}
  \caption{\label{interview-table} A summary of the research questions asked in the interviews.}
  \footnotesize{* Gartner AI Maturity model is briefly explained to the interviewee before being asked the question. \\
   Question is asked if it's valid and logical to be asked, thus interviews are semi-structured.}
\end{table*}

\section{Summary of key findings}
The key findings from the interviews are summarized in the table below. These insights provide a clear view of how AI is being integrated into Finnish health-tech SMEs, the challenges faced, and the opportunities for future growth are shown in Table~\ref{ai-key-findings}.

\begin{table*}
    \centering
    \begin{tabular}{lc}
        \hline
        \textbf{Category} & \textbf{Key Findings} \\
        \hline
        \textbf{AI in Products \& Services} & Used for analysis, potential future solutions, core AI products. \\
        \hline
        \textbf{Definition of AI} & Seen as a learning algorithm, quality-of-life enhancer, anomaly detector. \\
        \hline
        \textbf{AI Maturity Levels} & Companies mostly at levels 1, 2, and 3 of Gartner’s AI Maturity Model. \\
        \hline
        \textbf{AI Application Areas} & Mainly used for health analysis; some use in marketing and operational tools. \\
        \hline
        \textbf{Perceived Impact} & Indispensable for most companies; non-users acknowledge AI’s potential. \\
        \hline
        \textbf{Data Source} & Companies use in-house and external data sources, with privacy concerns. \\
        \hline
        \textbf{Computing Environment} & AI tools outsourced, private and public cloud services used. \\
        \hline
        \textbf{Challenges} & Strict regulations, market resistance, difficulty hiring AI talent. \\
        \hline
        \textbf{Benefits of AI} & Improved speed and accuracy, operational necessity. \\
        \hline
        \textbf{Wishes for 3rd Parties} & Lower-cost AI solutions, better access to data. \\
        \hline
        \textbf{Future Concerns} & Increased computational requirements, regulatory standardization. \\
        \hline
        \textbf{Advice for Companies} & Gather data, define problems, hire skilled professionals, experiment with AI. \\
        \hline
    \end{tabular}
    \caption{\label{ai-key-findings} Summary of key findings on AI adoption and challenges.}
\end{table*}
\end{document}